\documentclass[conference,10pt]{IEEEtran}
\usepackage{cite}
\usepackage{graphicx}
\usepackage{psfrag}
\usepackage{url}
\usepackage{amsmath}
\usepackage{array}
\usepackage{amssymb}
\usepackage{amsthm}
\usepackage{mathtools}
\usepackage{amsfonts}
\usepackage{graphicx}
\usepackage{epstopdf}
\usepackage{algorithm}
\usepackage{algorithmic}

\usepackage{setspace}
\usepackage{amscd}
\usepackage{mathrsfs}
\usepackage{epsfig}
\usepackage{color}
\usepackage{textcomp}
\usepackage{tikz}
\usepackage{comment}
\usetikzlibrary{shapes,snakes,plotmarks}

\usepackage{caption}
\usepackage{subcaption}

\IEEEoverridecommandlockouts

\DeclareMathOperator{\trace}{Tr}
\DeclareMathOperator*{\argmax}{argmax}
\DeclareMathOperator*{\argmin}{argmin}

\makeatletter
\newcommand\fs@spaceruled{\def\@fs@cfont{\bfseries}\let\@fs@capt\floatc@ruled
  \def\@fs@pre{\vspace{0.5\baselineskip}\hrule height.7pt depth0pt \kern2pt}%
  \def\@fs@post{\kern2pt\hrule\relax}%
  \def\@fs@mid{\kern2pt\hrule\kern2pt}%
  \let\@fs@iftopcapt\iftrue}
\makeatother

\title{BeamSync: Over-The-Air Carrier Synchronization in Distributed RadioWeaves}

\begin{document}
\author{\IEEEauthorblockN{Unnikrishnan Kunnath Ganesan, Rimalapudi Sarvendranath  and Erik G. Larsson}
		\IEEEauthorblockA{Department of Electrical Engineering (ISY)\\
		Link\"oping University, Link\"oping, SE-581 83, Sweden.\\
		Emails: \{unnikrishnan.kunnath.ganesan, sarvendranath.rimalapudi, erik.g.larsson\}@liu.se}
\thanks{This work was funded by the REINDEER project of the European Union`s Horizon 2020 research and innovation programme under grant agreement No. 101013425.}	
}
\maketitle
\thispagestyle{empty}

\begin{abstract}
In a distributed multi-antenna system, multiple geographically separated transmit nodes communicate simultaneously to a receive node. Synchronization of these nodes is essential to achieve a good performance at the receiver. RadioWeaves is a new paradigm of cell-free massive MIMO array deployment using distributed multi-antenna panels in indoor environments. In this paper, we study the carrier frequency synchronization problem in distributed RadioWeave panels. We propose a novel, over-the-air synchronization protocol, which we call as BeamSync, to synchronize all the different multi-antenna transmit panels. We also show that beamforming the synchronization signal in the dominant direction of the channel between the panels is optimal and the synchronization performance is significantly better than traditional beamforming techniques.
\end{abstract}


\section{Introduction}
\label{sec:Introduction}

Massive MIMO introduced as a wild academic idea in~\cite{marzetta2010noncooperative}, has become an integral part of the 5G wireless standard and is envisioned as a key technology for beyond 5G systems. It is shown to improve the spectral efficiency of the wireless networks~\cite{marzetta2016fundamentals}. To further improve the spectral efficiency, cell-free massive MIMO, which reduces the inter-cell interference by employing distributed access points (APs) with no cell boundaries, is introduced in~\cite{ngo2017cell}. Different architectures are studied with this cell-free networking approach. Recently, a new paradigm of cell-free network deployment, known as RadioWeaves is introduced in~\cite{van2019radioweaves}, in which distributed radio and compute resources are weaved into large surface areas such as walls. This new RadioWeaves infrastructure is envisioned to provide high connection density, reliability, and low latency at unprecedented energy efficiency. RadioWeaves can achieve high data rates with very low power consumption was shown in~\cite{ganesan2020radioweaves}.

Coherent reception of the signal at the user, which requires synchronization among the distributed transmitters, is critical to achieve the benefits of distributed architectures such as RadioWeaves. However, in practice, achieving synchronization is a challenging problem. Each transceiver in a communication system is equipped with a local oscillator circuit that generates carrier frequency based on a reference crystal oscillator. Due to mismatches in the reference oscillator circuits, different transceivers generate different carrier frequencies. Furthermore, the generated frequencies drift over time for instance, due to fluctuations in temperature and voltage. Hence, the carrier frequency at different transceivers will be different. This results in a carrier frequency offset between any two transceiver nodes, which degrades the performance of the communication system. In order to avoid this problem and to achieve carrier frequency synchronization in the global system for mobile (GSM) systems, frequency correction burst signals (FBs) are sent periodically through the frequency correction channel (FCCH)~\cite{gsmetsi}. After listening to FBs, receivers tune their local oscillators to match their carrier frequency with the transmitter. Different carrier frequency synchronization techniques were studied for a point-to-point orthogonal frequency division multiplexing (OFDM) system in~\cite{moose1994technique,schmidl1997robust,huang2006carrier}. 

The synchronization techniques developed for a point-to-point communication system  do not extend directly to a distributed communication system. 
This is because the receiver observes a combined signal from different transmit nodes. One possible way to achieve carrier frequency synchronization is to provide a common carrier frequency to these distributed transmit nodes through a wired fronthaul network. However, this is not a scalable solution as the number of distributed transmitters increase. To address this issue, over-the-air carrier synchronization methods were studied in~\cite{abari2015airshare,balan2013airsync,rogalin2014scalable}. AirShare technique proposed in~\cite{abari2015airshare}, uses a dedicated emitter to transmit two low-frequency tones over the air. The distributed transceivers use a dedicated circuit to receive these tones and generate their reference signal with the frequency equal to the difference of the two tones. This technique is robust to variations in temperature and supply voltage at the emitter. However, it uses out of band frequency resources. In AirSync technique studied in~\cite{balan2013airsync}, a primary AP transmits pilots continuously in the out of the data transmission band. The secondary APs receive  these pilots to estimate the frequency offset. This technique requires continuous transmission of the pilots from the primary AP and one dedicated receive antenna at each secondary AP. A pilot signaling between anchor APs, which form a connected cover of the network, in a special synchronization slot to estimate the frequency offset is proposed in~\cite{rogalin2014scalable}. These estimates are exchanged through a wired fronthaul connecting the distributed transmit nodes. The scheme requires geographically dispersed anchor nodes and requires high anchor AP density.

{\em Focus and Contributions:}
In this paper, we study over-the-air carrier frequency synchronization in distributed RadioWeaves array deployment. The contributions of the paper can be summarized as follows:
\begin{itemize}
\item We propose a novel, over-the-air, carrier frequency synchronization protocol based on digital beamforming, which we shall refer to as BeamSync, for distributed multi-antenna panels in RadioWeaves infrastructure. In BeamSync, we consider one of the panels as primary panel and others as secondary panels, which need to synchronize with the primary panel. BeamSync removes the requirement of dedicated circuits for synchronization at transceiver nodes unlike~\cite{abari2015airshare,balan2013airsync}. Moreover, the scheme does not exchange calibration data through wired fronthaul connections and enables a faster carrier frequency synchronization.

\item BeamSync exploits the diversity benefits of the multiple antennas at each panel to beamform the synchronization signal. The primary panel beamforms the frequency synchronization signal towards the secondary panels in the dominant direction of the channel between the primary and secondary panels. The secondary panels estimate their frequency offset with respect to primary using signal processing techniques. We show that the optimal beamforming direction which minimizes the offset estimation error is the dominant direction of the channel between the panels in which the signal is received.

\item Our simulations show that BeamSync performs significantly better than analog beamforming. We show this for both Rayleigh fading channels with omni directional antennas and line-of-sight channels with directional patch antennas. The results also show that the estimation error decreases significantly as the number of antennas at the panels' increases.
\end{itemize}

\textbf{\textit{Notations:}} Bold, lowercase letters are used to denote vectors and bold, uppercase letters are used to denote matrices. $\mathbb{R}$ and $\mathbb{C}$ denote the set of real and complex numbers respectively. For a matrix $\mathbf{A}$, $\mathbf{A}^*$, $\mathbf{A}^\text{T}$ and $\mathbf{A}^\text{H}$ denotes conjugate, transpose and conjugate transpose of the matrix $\mathbf{A}$ respectively. $\mathcal{CN}(0,\sigma^2)$ denotes a circularly symmetric complex Gaussian random variable with zero mean and variance equal to $\sigma^2$. Identity matrix is of size $ K $ denoted by $ \mathbf{I}_K $.

\section{System Model}
\label{sec:SystemModel}

\begin{figure}[t]
	\centering
	\includegraphics[scale=0.25]{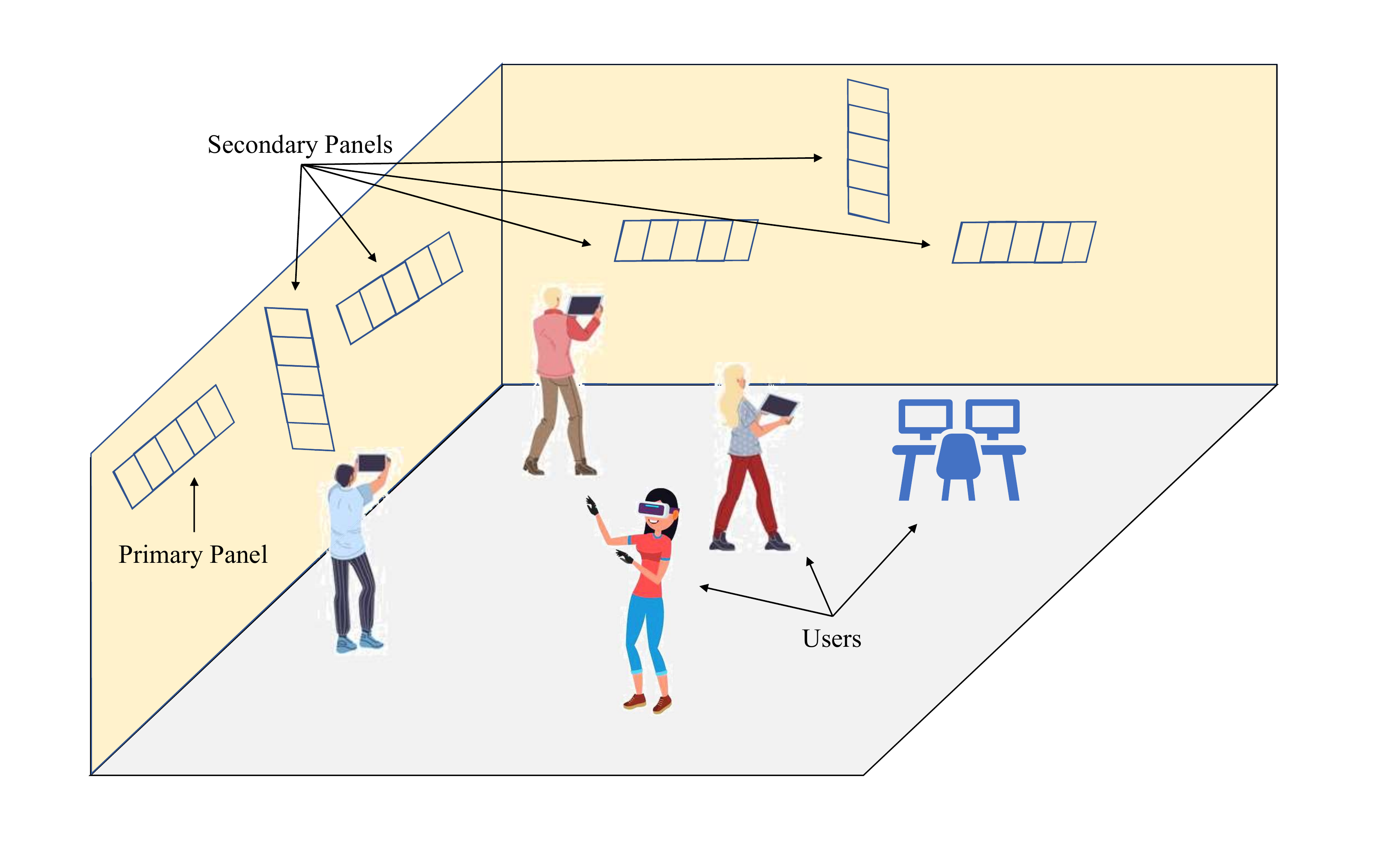}
	\caption{Distributed RadioWeaves array deployment system model.}
	\label{fig:SystemModel}
	\vspace{-10pt}
\end{figure}

We consider a distributed RadioWeaves array deployment, which consists of multiple  geographically separated panels communicating simultaneously to the users as shown in Fig.~\ref{fig:SystemModel}. Each panel is equipped with multiple antennas, each with its own radio frequency (RF) chain,  and one local oscillator circuit to generate the carrier frequency.  There is no mismatch between the carrier frequency among RF chains in the same panel, as all of them are driven by the same oscillator circuit of the panel. However, the carrier frequencies generated at different panels will differ. Hence, to synchronize the carrier frequency among the panels to a common reference, we nominate one of the panels as the primary panel and consider its carrier frequency as the reference. The remaining panels, which we refer to as secondary panels, synchronize with the primary. 

Let $ N_s $ denote the number of secondary panels. Let $ M_p $ denote the number of antennas at the primary panel and $ M_{s,i} $ denote the number of antennas at the $ i^\text{th} $ secondary panel, for $i\in\{1,2,\ldots,N_{s}\}$. Let $\mathbf{G}_i \in \mathbb{C}^{M_p\times M_{s,i}}$ denote the complex channel gain matrix between the primary panel and the $ i^\text{th} $ secondary panel. We assume that the channel is reciprocal. Let $f_p$ and $f_{s,i}$ denote the carrier frequencies at the primary panel and the $ i^\text{th} $ secondary panel, respectively. Then, the carrier frequency offset of the $ i^\text{th} $ secondary panel with respect to the primary panel is given by $\Delta_i = f_p - f_{s,i}$. Each of the secondary panels estimates its $\Delta_i$ and compensates during data transmission to the users.

\section{Carrier Synchronization}
\label{sec:CarrierSync}
In this section, we first describe our BeamSync protocol based on beamforming for one secondary panel. We then show that the optimal beamforming direction is the dominant direction of the channel between the primary and secondary panel in which the signal is received. We then propose a method to find the dominant  beamforming direction in BeamSync. Finally, we generalize the BeamSync protocol for multiple secondary panels.

\subsection{BeamSync Protocol} \label{sec:SyncProtocol}
We now describe the BeamSync protocol for one secondary panel. For simplicity, we drop the index $ i $ of the secondary panel in the notations. Therefore, $ M_s $, $\mathbf{G}\in\mathbb{C}^{M_p\times M_s}$, and~$\Delta$, denote the number of antennas at the secondary panel, the channel matrix between the primary and secondary panel, and the frequency offset, respectively. The protocol consists of two stages described as follows:

\subsubsection*{Stage-I}
The secondary panel transmits a orthonormal pilot sequence of length $ \tau_p \geq M_s $ from each of its antennas. Let the columns of the matrix $\mathbf{\Phi} \in \mathbb{C}^{\tau_p \times M_s}$, where  $\mathbf{\Phi}^\text{H}\mathbf{\Phi}=\mathbf{I}_{\tau_p}$,  denote the set of orthonormal pilot sequences. Let $\phi(n)$, denote the $ n^\text{th} $ row of $\mathbf{\Phi}$. Thus, at the $n^{\text{th}}$ time instant, the signal received at the primary panel $\mathbf{y}_p \in \mathbb{C}^{M_p \times 1} $  can be expressed as
\begin{equation}
\mathbf{y}_p (n) = \sqrt{\rho} \mathbf{G} \phi^\text{H}(n) e^{j 2 \pi n \Delta} + \mathbf{w}_p(n), 
\end{equation}
where $\rho$ is the normalized signal to noise ratio (SNR) and $\mathbf{w}_p(n) \in \mathbb{C}^{M_p \times 1} $ is the additive noise with each of the elements independent and identically distributed (i.i.d.) $\mathcal{CN}(0,1)$. 
Let 
\begin{equation}
\mathbf{D}_{\Delta,\tau} = \text{diag}\{e^{j2\pi \Delta},e^{j2\pi 2 \Delta},\dots,e^{j2\pi \tau \Delta }\} \in \mathbb{C}^{\tau\times \tau}.
\end{equation}
The collective signal received in $\tau_p$ time instants at the primary panel, $\mathbf{Y}_p~=~[ \mathbf{y}_p(1) \ \mathbf{y}_p(2) \ \cdots \ \mathbf{y}_p (\tau_p)]  $, can be written as 
\begin{equation}
	\mathbf{Y}_p = \sqrt{\rho} \mathbf{G\Phi}^{\text{H}} \mathbf{D}_{\Delta,\tau_p}  + \mathbf{W}_p,  
\end{equation}
where $\mathbf{W}_p = [\mathbf{w}_p(1) \ \mathbf{w}_p(2) \ \cdots \ \mathbf{w}_p(\tau_p) ]$.

\subsubsection*{Stage-II}
The primary panel processes the signal $\mathbf{Y}_p$ received in stage-I and determines a beamforming vector $\mathbf{a}~\in~\mathbb{C}^{M_p\times1}$ such that $ \lVert \mathbf{a} \rVert $=1. It then beamforms a $ N $ length frequency synchronization signal $\mathbf{x}$. The received signal at the secondary, $\mathbf{y}_s (n)\in \mathbb{C}^{M_s \times 1} $ at the $n^{\text{th}}$ time instant is given by 
\begin{equation}
	\mathbf{y}_s (n) = \sqrt{\rho} \mathbf{G}^\text{T} \mathbf{a} x(n) e^{-j 2 \pi n \Delta} + \mathbf{w}_s(n),
\end{equation}
where  $x(n)$ is the $n^\text{th}$ component of signal $\mathbf{x}$ and $\mathbf{w}_s(n)~\in~\mathbb{C}^{M_s \times 1} $ is the additive noise with i.i.d. $\mathcal{CN}(0,1)$ entries. The collective signal received over $ N $ times instants at secondary panel,  $\mathbf{Y}_s = [ \mathbf{y}_s(1) \ \mathbf{y}_s(2) \ \cdots \ \mathbf{y}_s (N)] $ can be written as 
\begin{equation}
	\label{eqn:SecondarySyncSignal}
	\mathbf{Y}_s = \sqrt{\rho} \mathbf{G}^{\text{T}} \mathbf{ax}^\text{T} \mathbf{D}^*_{\Delta,N}  + \mathbf{W}_s,
\end{equation}
where $\mathbf{W}_s = [\mathbf{w}_s(1) \ \mathbf{w}_s(2) \ \cdots \ \mathbf{w}_s(N) ]$. Secondary panel needs to estimate its frequency offset $\Delta $ with respect to the primary from (\ref{eqn:SecondarySyncSignal}). The channel $\mathbf{G}$ and the beamforming vector $\mathbf{a}$ are unknown at the secondary panel. Let $\mathbf{b} = \mathbf{G}^{\text{T}} \mathbf{a}$ denote the effective channel. Then (\ref{eqn:SecondarySyncSignal}) can be rewritten as 
\begin{equation}
	\label{eqn:SecondarySyncSignal_2}
	\mathbf{Y}_s = \sqrt{\rho} \mathbf{bx}^\text{T} \mathbf{D}^*_{\Delta,N}  + \mathbf{W}_s. 
\end{equation}
The joint maximum likelihood estimates of $\mathbf{b}$ and $\Delta$ are given by 
\begin{equation}
	\label{eqn:joint_estimate}
(\hat{\mathbf{b}},\hat{\Delta}) = \argmin_{\mathbf{b},\Delta} \lVert \mathbf{Y}_s - \sqrt{\rho} \mathbf{bx}^\text{T} \mathbf{D}^*_{\Delta,N} \rVert^2.
\end{equation}
Solving (\ref{eqn:joint_estimate}) using non-linear least squares estimation in Gaussian noise~\cite[Sec. 8.9]{kay1993fundamentals} with $\mathbf{b}$ as a nuisance parameter, the estimates of $\mathbf{b}$ and $\Delta$ are given by 
\begingroup
\allowdisplaybreaks
\begin{align}
\hat{\mathbf{b}} & = \frac{\mathbf{Y}_s \mathbf{D}_{\Delta,N} \mathbf{x}^*}{ \sqrt{\rho} \lVert \mathbf{x} \rVert ^2 },  \\
\hat{\Delta} & = \argmax_\Delta \lVert \mathbf{Y}_s\mathbf{D}_{\Delta,N} \mathbf{x}^* \rVert^2.
\end{align}
\endgroup
The secondary panel uses $\hat{\Delta}$ to derotate its transmitted signals to synchronize with the primary panel. 

\subsection{Optimal Beamforming Direction}
\label{sec:OptimalBeamforming}
In this section, we derive the optimal beamforming direction that minimizes the offset estimation error. We look at the conditions for which the Cram\'er Rao Bound (CRB) on the estimate of $\Delta$, is minimized.

Let $ (\cdot)_\text{R} $ and $ (\cdot)_\text{I} $ denote the real and imaginary parts of a complex number, respectively. Then $\mathbf{b} = \mathbf{b}_\text{R} + j \mathbf{b}_\text{I}$ and $ \mathbf{y}_s(n) = \mathbf{y}_{s\text{R}}(n) + j\mathbf{y}_{s\text{I}}(n) $. Let 
\begin{equation}
\boldsymbol{\theta} = [ \mathbf{b}_\text{R}^\text{T} \ \mathbf{b}_\text{I}^\text{T} \ \Delta]^\text{T},
\end{equation}
be the unknown parameter at the secondary panel. From (\ref{eqn:SecondarySyncSignal_2}), the signal received at the $ n^\text{th} $ time instant, $\mathbf{y}_s(n)$ is distributed as $\mathcal{CN}(\sqrt{\rho}\mathbf{b}x(n)e^{-j 2 \pi \Delta n}, \mathbf{I})$. We assume that the frequency synchronization signal $\mathbf{x}$ is real-valued. Thus, $\bar{\mathbf{y}}_s(n)~=~[\mathbf{y}_{s\text{R}}^\text{T}(n) \ \mathbf{y}_{s\text{I}}^\text{T}]^\text{T}(n)~\in~\mathbb{R}^{2M_s\times 1}$ is distributed as $\mathcal{N}(\boldsymbol{\mu}_n(\boldsymbol{\theta}),\mathbf{C(\boldsymbol{\theta})})$, where $ \boldsymbol{\mu}_n(\boldsymbol{\theta}) $ and $\mathbf{C}(\boldsymbol{\theta})$ denote the mean and covariance of $\bar{\mathbf{y}}_s$, respectively parameterized by $ \boldsymbol{\theta} $, and are given by
\begin{equation}\label{eqn:mean}
\boldsymbol{\mu}(\boldsymbol{\theta}) = \sqrt{\rho} x(n) \begin{bmatrix*}[r]
\mathbf{b}_\text{R} \cos(2\pi n \Delta) + \mathbf{b}_\text{I} \sin(2\pi n \Delta) \\
-\mathbf{b}_\text{R} \sin(2\pi n \Delta) + \mathbf{b}_\text{I} \cos(2\pi n \Delta) \\
\end{bmatrix*},
\end{equation}
\begin{equation}\label{eqn:covariance}
\mathbf{C}(\boldsymbol{\theta})  = \frac{1}{2}\mathbf{I}_{2M_s}. 
\end{equation}
Using Slepian Bang theorem \cite[Sec. 3.9]{kay1993fundamentals}, each element of the Fisher information matrix (FIM) of $\boldsymbol{\theta}$ at $ n^\text{th} $ time instant, $ \mathbf{J}_n(\boldsymbol{\theta})~\in~\mathbb{R}^{(2M_s+1) \times (2M_s+1)} $, can be computed as 
\begin{equation}
	[\mathbf{J}_n(\boldsymbol{\theta})]_{k,l} = 
	\begin{aligned}
	& \left[\frac{\partial \boldsymbol{\mu}(\boldsymbol{\theta})}{\partial \boldsymbol{\theta}_k}\right]^\text{T} \mathbf{C}^{-1}(\boldsymbol{\theta})  \left[\frac{\partial \boldsymbol{\mu}(\boldsymbol{\theta})}{\partial \boldsymbol{\theta}_l}\right] ~~~  \\
	& + \frac{1}{2} \trace\left[ \mathbf{C}^{-1}(\boldsymbol{\theta}) \frac{\partial \mathbf{C}(\boldsymbol{\theta})}{\partial \boldsymbol{\theta}_k}\mathbf{C}^{-1}(\boldsymbol{\theta}) \frac{\partial \mathbf{C}(\boldsymbol{\theta})}{\partial \boldsymbol{\theta}_l}  \right] , 
	\end{aligned}
\end{equation}
where $ [\cdot]_{k,l} $ denotes the $ (k,l)^\text{th} $ element. By computing the partial derivatives of (\ref{eqn:mean}) and (\ref{eqn:covariance}), we obtain
\begin{equation}\label{eqn:FIM_n}
	\mathbf{J}_n(\boldsymbol{\theta}) = 2 \rho x^2(n) 
	\begin{bmatrix*}[c]
		 \mathbf{I}_{M_s} & \mathbf{0} & 2\pi n \mathbf{b}_{\text{I}}\\
 		\mathbf{0} &  \mathbf{I}_{M_s} & -2\pi n  \mathbf{b}_{\text{R}} \\
		2\pi n  \mathbf{b}_{\text{I}}^{\text{T}} & -2\pi n \mathbf{b}_{\text{R}}^{\text{T}} & 4\pi^2 n^2  \lVert \mathbf{b} \rVert^2
	\end{bmatrix*}.
\end{equation}

The received signal $\mathbf{y}_s(n)$ is independent for different time instants. Thus, using the additive property of FIM, the overall FIM of $\boldsymbol{\theta}$, $ \mathbf{J}(\boldsymbol{\theta}) $, is given by 
\begin{equation}\label{eqn:FIM}
	\mathbf{J}(\boldsymbol{\theta}) = \sum_{n=1}^{N}    \mathbf{J}_n(\boldsymbol{\theta}).
\end{equation}
The CRB of $\hat{\Delta}$ can be computed from $ \text{J} (\boldsymbol{\theta}) $ as 
\begin{equation}
	\text{CRB}(\hat{\Delta}) = \left[\mathbf{J}^{-1}(\boldsymbol{\theta}) \right]_{2M_s+1,2M_s+1}, 
\end{equation}
which is the lower right corner element of $ \mathbf{J}^{-1}(\boldsymbol{\theta}) $. Using the inverse of a block partitioned matrix~\cite[Sec. 0.7.3]{horn2012matrix}, the CRB of $\hat{\Delta}$ is given by 
\begin{equation}\label{eqn:CRB_Delta}
		\text{CRB}(\hat{\Delta}) = \frac{1}{ 8 \pi^2 \rho \lVert \mathbf{b} \rVert^2 \left( \sum_{n=1}^N n^2x^2(n) - \frac{(\sum_{n=1}^N n x^2(n))^2}{\sum_{n=1}^N x^2(n)}\right)  }.
\end{equation}

From (\ref{eqn:CRB_Delta}), the CRB of $\hat{\Delta}$ will be minimized when $ \lVert\mathbf{b} \rVert^2~=~\lVert \mathbf{G}^\text{T}\mathbf{a} \rVert^2 $ is maximized. Let the singular value decomposition (SVD) of the channel $\mathbf{G}$ be 
\begin{equation}
		\mathbf{G} = \mathbf{U\Sigma V}^{\text{H}}, 
		\label{eqn:SVD_channel}
\end{equation}
where $\mathbf{U} \in \mathbb{C}^{M_p\times M_p} $ and $\mathbf{V} \in \mathbb{C}^{\tau_p\times \tau_p} $ are unitary matrices and $\mathbf{\Sigma}\in\mathbb{R}^{M_p\times \tau_p}$ is a diagonal matrix with singular values of $\mathbf{G} $ in decreasing order. Then, 
\begin{equation}\label{eqn:OptimalBeamForming}
	\addtocounter{equation}{-1}
	\begin{subequations}
		\begin{aligned}
			\lVert \mathbf{G}^\text{T}\mathbf{a}\rVert ^2 & = \mathbf{a}^\text{H}\mathbf{G}^\text{*} \mathbf{G}^\text{T}  \mathbf{a} \\
			&= \mathbf{a}^\text{H}\mathbf{U}^\text{*}\mathbf{\Sigma} \mathbf{\Sigma}^\text{T} \mathbf{U}^\text{T} \mathbf{a}.
		\end{aligned}
	\end{subequations}
\end{equation}
From (\ref{eqn:OptimalBeamForming}), $ \lVert \mathbf{G}^\text{T}\mathbf{a}\rVert ^2 $ is the Rayleigh quotient of matrix $\mathbf{G}^*\mathbf{G}^\text{T}$ with vector $\mathbf{a}$ and can be maximized by choosing $ \mathbf{a}~=~\mathbf{u}_1^*$. The vector $\mathbf{u}_1$, is the first column of matrix $\mathbf{U}$. Hence, the optimal beamforming direction $\mathbf{a}$, corresponds to the dominant direction of the channel in which the signal will be received at the secondary panel. 

From (\ref{eqn:CRB_Delta}), for estimating $\Delta$, the synchronization signal length $ N $ should be at least 2. Moreover, from (\ref{eqn:CRB_Delta}), the frequency offset estimate $\hat{\Delta}$ can be improved by increasing the SNR $\rho$, as well as increasing the synchronization sequence length $ N $. 

\begin{figure}[t]
	\centering
	\includegraphics[scale=0.4]{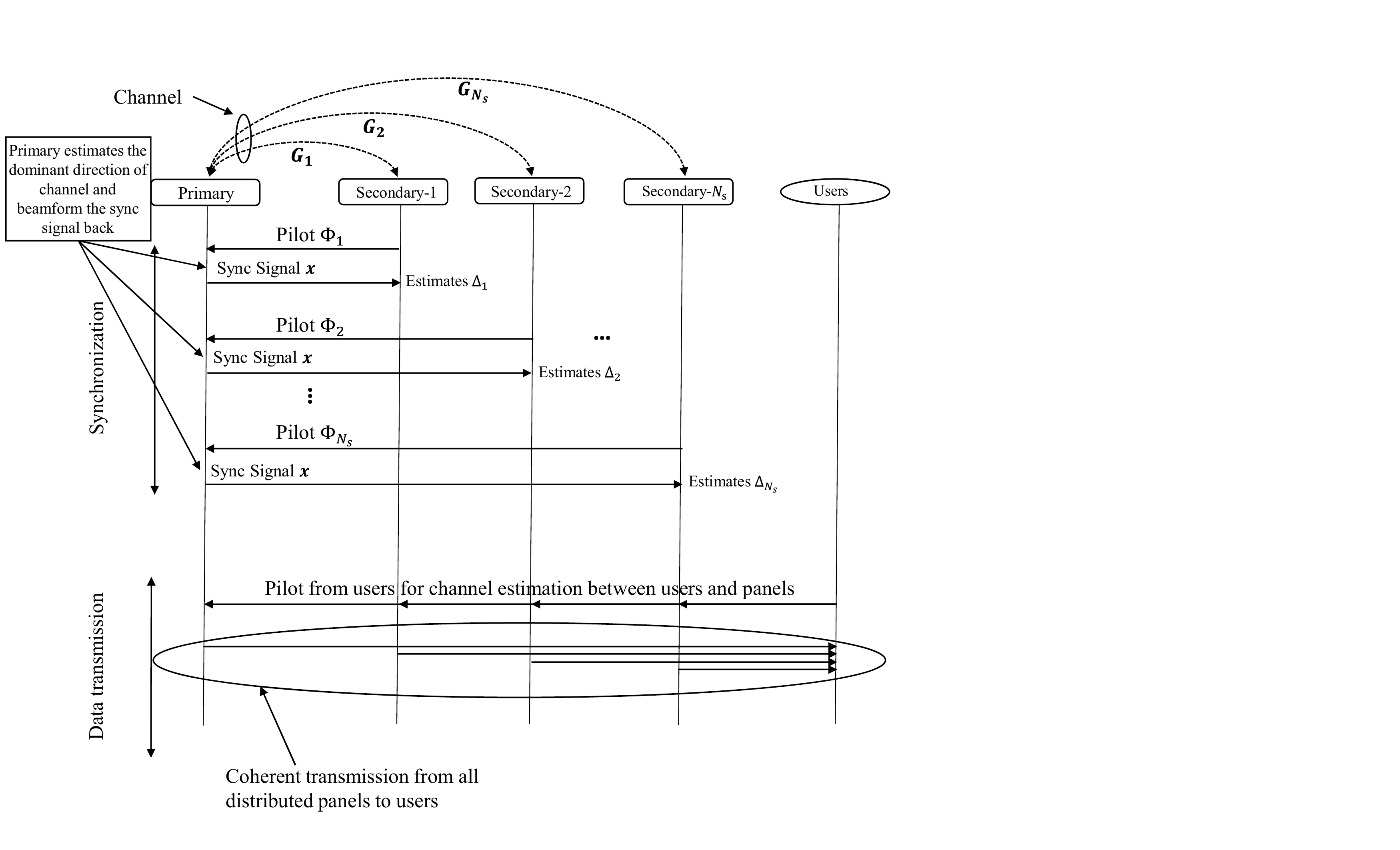}
	\caption{BeamSync protocol.}
	\label{fig:SyncProtocol}
\end{figure}

\subsection{Estimating Beamforming Direction in BeamSync}
From Sec. \ref{sec:OptimalBeamforming}, it is evident that the optimal direction to beamform the synchronization signal is the dominant direction of the channel in which the secondary panel receives the signal from the primary panel. In practice, the channel $\mathbf{G}$ is not perfectly known at the primary panel. However, as the channel is reciprocal, the primary panel can estimate the dominant direction from the signal received from the secondary panel, without the need to estimate the channel. The primary panel listens to the pilot signal $\mathbf{\Phi}$ which is transmitted in all directions by the secondary panel in stage-I of synchronization protocol and computes the dominant direction in which the signal was received. It can be mathematically expressed as SVD of $\mathbf{Y}_p$ given by 
\begin{equation}
	\mathbf{Y}_p = \mathbf{U}_p\mathbf{\Sigma}_p \mathbf{V}_p^{\text{H}}, 
	\label{eqn:SVD_computation}
\end{equation}
where $\mathbf{U}_p \in \mathbb{C}^{M_p\times M_p} $ and $\mathbf{V}_p \in \mathbb{C}^{\tau_p\times \tau_p} $ are unitary matrices and $\mathbf{\Sigma}_p\in\mathbb{R}^{M_p\times \tau_p}$ is a diagonal matrix with singular values of $\mathbf{Y}_p $ in decreasing order. The columns of $\mathbf{U}_p$ corresponds to the direction of the received signal ordered according to the dominance of power received in each direction. Hence the optimal beamforming direction is given by $\mathbf{a}=\mathbf{u}_{p1}^*$, where $\mathbf{u}_{p1}$ is the first column of $\mathbf{U}_p$. As SNR increases, primary panel will be able to perfectly determine the dominant direction of the channel asymptotically, i.e., $ \mathbf{u}_{p1} \rightarrow \mathbf{u}_{1} $.

\subsection{Over-The-Air Carrier Synchronization Protocol}
We generalize the proposed BeamSync protocol for multiple secondary panels and the communication flow is shown in Fig.~\ref{fig:SyncProtocol}. During the cold start or initialization of the entire communication system, all the distributed transceivers will be out of sync. After the initial power up, all the secondary panels will synchronize with the primary panel in a sequential fashion using the BeamSync protocol. Afterwards, the distributed panels can start joint coherent transmission to the users in the data transmission phase. Moreover, as the carrier frequency synchronization is done over-the-air without the need of wired fronthaul connections, it enables a faster carrier frequency synchronization.

\begin{figure}[t]
	\centering
	\includegraphics[scale=0.23]{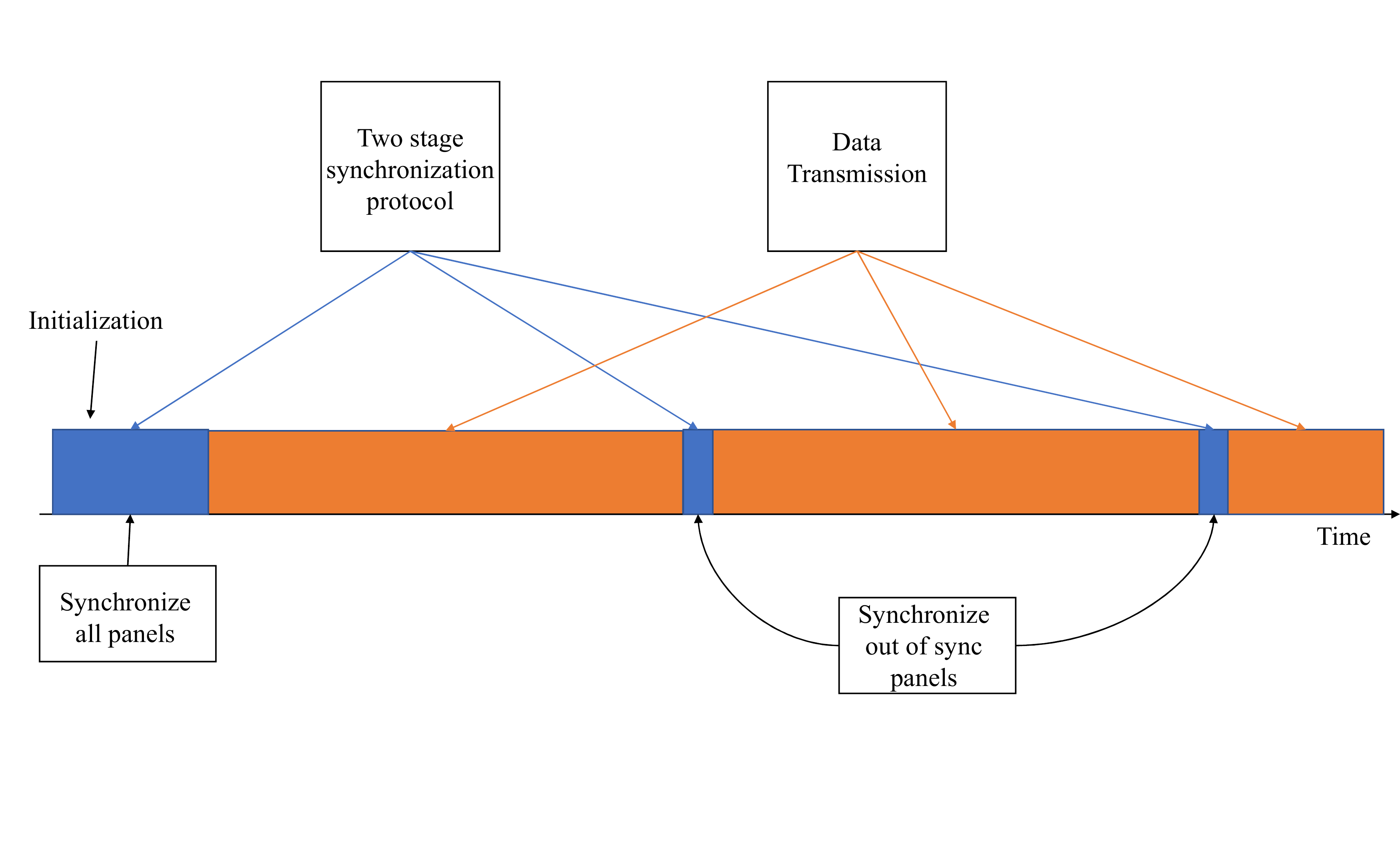}
	\caption{Synchronization procedure over time.}
	\label{fig:SyncProtocolOverTime}
\end{figure}

The frequency generated through the local oscillator circuit can drift over time, for instance due to fluctuations in temperature and voltage. This frequency drift is negligible in a coherence interval. Hence, after the cold start, the synchronization procedure needs to be done when the secondary panel goes out of sync, based on a need basis. Thus, the synchronization procedure dispersed over time is represented in Fig.~\ref{fig:SyncProtocolOverTime}.

\section{Simulations}
\label{sec:sSimulations}
In this section, we first outline the simulation parameters. We then outline other schemes for performance benchmarking our proposed schemes. Finally, the results of our study on carrier synchronization in distributed RadioWeaves array deployment is discussed.

{\em Simulation Parameters: }
For simulations, we consider the number of antennas at the primary panel and secondary panel as $ M_p = M_s = 16 $. The pilot signal length $\tau_p = M_s$.  The length of the synchronization signal transmitted from the primary panel, $ N=100 $. The Monte Carlo trials considered is $ 10^5 $. We consider the frequency synchronization signal $\mathbf{x}$ as
\begin{equation}\label{eqn:syncsignal}
	\mathbf{x} = [1 \ \sin(2\pi f) \ \sin(4\pi f) \cdots \ \sin(2\pi f (N-1) )]^\text{T},
\end{equation}
where $ f $ is the frequency and is chosen to have four full cycles of sinusoid signal in $ N $ time instants. 

\subsection{Performance Benchmarking Schemes}
In our proposed scheme, the beamforming vector can be pointed in any direction in 3-dimensional environment and can be done by digital signal processing techniques. Hence, BeamSync is a fully digital beamforming scheme. For comparison in the figures we refer our proposed schemes as follows:
\begin{enumerate}
\item {\em BeamSync:} Proposed scheme, where the received signal at primary is used to find the beamforming direction $\mathbf{a}=\mathbf{u}_{p1}^*$.
\item {\em BeamSync-Genie:} Proposed scheme, where through an aid of a genie, we consider that the primary panel perfectly knows the channel $\mathbf{G}$. The beamforming direction is $\mathbf{a}=\mathbf{u}_1^*$.
\end{enumerate}

\begin{figure}[t]
	\centering
	\includegraphics[scale=0.45]{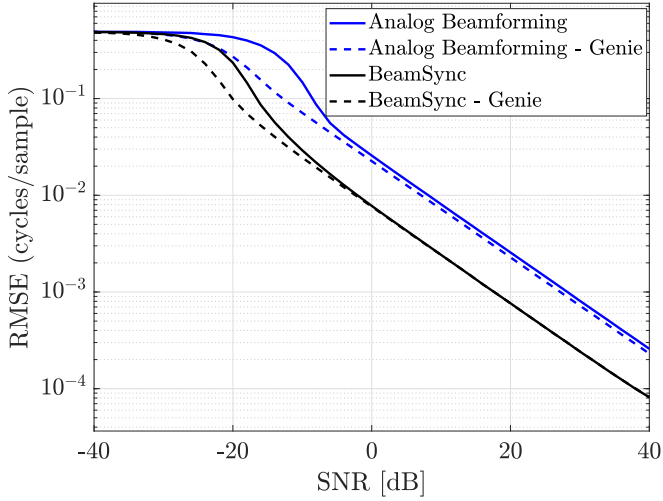}
	\caption{RMSE in Rayleigh fading Channel}
	\label{fig:PerfRayleigh}
\end{figure}

We compare our proposed schemes with the following beamforming techniques: 
\begin{enumerate}
\setcounter{enumi}{2}

\item \label{fixedgrid}{\em Analog beamforming:} In this scheme, the primary panel performs transmit beamforming and secondary panel performs receive beamforming. The beamforming vectors at both the panels are chosen from a fixed set of beams. In our numerical example, we consider columns of a DFT matrix as the possible set of orthogonal beams. Let $ \{ \mathbf{f}_{p,k} \in\mathbb{C}^{M_p\times 1}, ~k=1,2,\cdots,M_p  \} $    and $ \{ \mathbf{f}_{s,l} \in\mathbb{C}^{M_s\times 1}, ~l=1,2,\cdots,M_s \}$ be the fixed set of beams available at the primary and secondary panels, respectively. The transmit and receive beamforming vectors are chosen such that the received signal power is maximized. 
Let 
\begin{equation}
k = \argmax_{ k'} \ \lVert \mathbf{f}_{p,k'}^\text{H}\mathbf{Y}_p \rVert ^2, \ l = \argmax_{ l'} \ \lVert \mathbf{f}_{s,l'}^\text{H}\mathbf{Y}_s \rVert ^2.
\end{equation}

Then, the transmit beamforming vector is $\mathbf{a}_p~=~\mathbf{f}_{p,k}^* $, and the receive beamforming vector is $\mathbf{a}_s~=~\mathbf{f}_{s,l}$. 

\item {\em Analog beamforming-Genie:} Same as \ref{fixedgrid}), but we choose the beamforming vectors based on the perfectly known channel $\mathbf{G}$ through a genie at both primary and secondary panels. Let 
\begin{equation}
	(k,l) = \argmax_{ k',l' } \ \lvert \mathbf{f}_{p,k'}^\text{H} \mathbf{G} \mathbf{f}_{s,l'} \rvert^2.
\end{equation}
Then, the transmit beamforming vector is $\mathbf{a}_p~=~\mathbf{f}_{p,k}^* $, and the receive beamforming vector is $\mathbf{a}_s~=~\mathbf{f}_{s,l}$. 
\end{enumerate}

\subsection{Results}

\begin{figure}[t]
	\centering
	\includegraphics[scale=0.45]{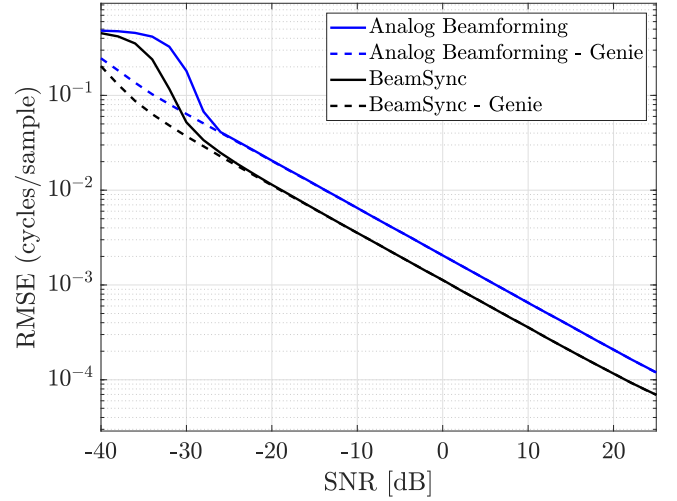}
	\caption{RMSE in direct line of sight channel}
	\label{fig:PerfLos}
\end{figure}

First, we consider a Rayleigh fading channel between the primary and the secondary panels. Thus, each element in $\mathbf{G}$ is i.i.d. $\mathcal{CN}(0,1)$. We consider the antennas to be omni-directional such that the signal can be transmitted and received in all directions. We use the root mean square error (RMSE) of the frequency offset estimate as to the performance metric for comparison. The performance of different schemes in the Rayleigh fading scenario is shown in Fig.~\ref{fig:PerfRayleigh}. From the plot, it can be seen that as the SNR increases, the RMSE decreases. When beamforming is done in the dominant direction determined from the perfect channel matrix, RMSE is lower for all SNR values among all the techniques. The performance of the proposed BeamSync protocol, which uses the dominant direction determined from the received vector, improves as SNR increases and matches with BeamSync-genie scheme at high SNR. This is because, as the SNR increases, the dominant direction chosen by BeamSync scheme becomes close to the one chosen from the perfect knowledge of the channel. Scheme 3, which uses analog beamforming, performs worse compared to other two schemes for all the SNR values. This shows that the fully digital beamforming in dominant direction yields significant performance gain compared to the analog beamforming with fixed beams. For example, for a fixed RMSE requirement, the SNR gain is approximately $ 10 $~dB for BeamSync.

\begin{figure}[t]
	\centering
	\includegraphics[scale=0.45]{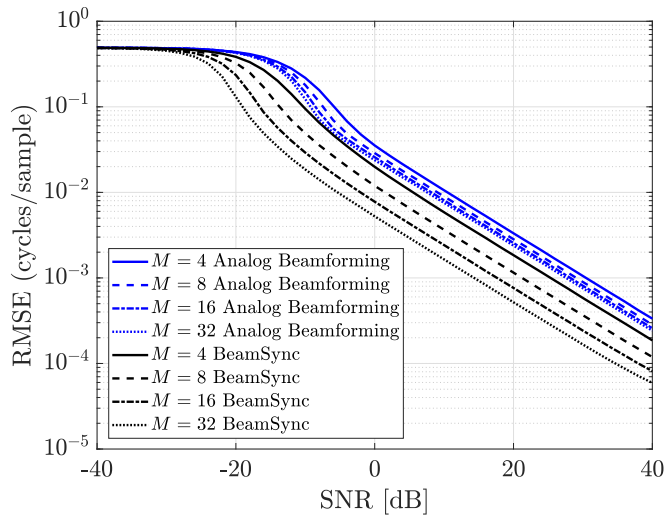}
	\caption{Performance of BeamSync with different number of antennas $ M_p=M_s=M $, pilot signal length $\tau_p=M_s$.}
	\label{fig:BeamSyncPerformance}
\end{figure}

Fig.~\ref{fig:PerfLos} compares the performance of the proposed synchronization schemes with other schemes for an indoor RadioWeaves deployment, where the channel between the panels will be dominated by the direct line of sight component~\cite{ganesan2020radioweaves}. We consider the panels are distributed in a $ 100 \text{m}\times100\text{m}\times10\text{m} $ room. We consider directional patch antennas on the panels, and the primary and secondary panels are on adjacent walls. The channel and antenna design parameters used are as in \cite{ganesan2020radioweaves}. Due to the strong line of sight signal, frequency offset can be better estimated at low SNR values compared to Rayleigh fading scenario. Similar to the Rayleigh fading case, BeamSync scheme matches with BeamSync-genie scheme at high SNR and performs better than the analog beamforming scheme. In this example, for a fixed RMSE requirement, the SNR gain is approximately $ 5 $~dB for BeamSync compared to the analog scheme. 

Fig.~\ref{fig:BeamSyncPerformance} compares the performance of the BeamSync protocol and analog scheme, when a different number of antennas are deployed at the panel. From the figure, it can be seen that for a fixed RMSE requirement, the SNR requirement reduces by $ 3 $~dB when the number of antennas is doubled at the panels for the BeamSync protocol. This is because the signal can be steered better in the desired direction as the number of antennas increases in BeamSync~\cite{marzetta2016fundamentals}. However, when the analog beamforming is used, the gain in performance is negligible as the number of antennas increases. 

\section{Conclusion}
\label{sec:Conclusion}
In this paper, we studied the carrier frequency synchronization in the distributed RadioWeaves array deployment. We proposed a novel, over-the-air carrier synchronization protocol, BeamSync, based on digital beamforming to synchronize different multi-antenna transmit panels in Radioweaves. We showed that sending the frequency synchronization signal burst in the dominant direction of the channel between the panels is optimal. We also proposed a scheme to estimate the beamforming direction without estimating the channel. We compared our scheme with analog beamforming scheme and showed that our proposed protocol can achieve better carrier frequency offset estimation. This is due to the improved SNR by beamforming and spatial processing gain. Moreover, the proposed protocol allows fast synchronization among the distributed panels. Also, we showed that, the better the synchronization signal burst is steered towards the secondary panel, the better is the offset estimation performance. 

\bibliographystyle{IEEEtran}
\bibliography{references}

\end{document}